\documentclass{pazhb_eng}
\usepackage[hyperfootnotes=false]{hyperref}
\usepackage{color}
\usepackage{epsf}
\usepackage[utf8]{inputenc}
\usepackage[T2A]{fontenc}
\usepackage{amsmath}
\usepackage{amsfonts}
\usepackage{amssymb}
\usepackage{graphicx}
\usepackage{gensymb}
\usepackage{float}
\usepackage{longtable}
\usepackage{changes}
\usepackage{wrapfig}
\usepackage{multicol}
\usepackage{makecell}
\usepackage{natbib}
\usepackage{amsmath}
\usepackage{amssymb}
\usepackage{textcomp}   
\usepackage{ion}        

\begin{document}

\journalinfo{2025}{51}{10}{604}[619]
\UDK{///}
\setcounter{page}{1}
\title{\bf Study of Massive OBA Stars with X-ray Emission}
\author{
E.~A.~Nikolaeva\address{1}\email{evgeny.nikolaeva@gmail.com}, I.~F.~Bikmaev\address{1,2},
E.~N.~Irtuganov\address{1,2}, M.~R.~Gilfanov\address{3,4}, R.~A.~Sunyaev\address{3,4}, P.~S.~Medvedev\address{3}
\addresstext{1}{Kazan (Volga Region) Federal University, Kremlevskaya 18, Kazan 420008, Russia}
\addresstext{2}{Tatarstan Academy of Sciences, Bauman 20, Kazan 420111, Russia}
\addresstext{3}{Space Research Institute, Russian Academy of Sciences, Profsoyuznaya 84/32, 117997 Moscow, Russia}
\addresstext{4}{Max Planck Institute for Astrophysics, Karl-Schwarzschild-Strasse 1, Garching bei München D-85741, Germany}
}

\shortauthor{E.A. Nikolaeva et al.}

\shorttitle{Study of Massive OBA Stars with X-ray Emission}

\begin{abstract}
A study of the physical parameters of a sample of 15 OBA-type stars with detected X-ray emission from the Spektr-RG/eROSITA telescope is presented. While X-ray emission from cool stars (spectral types F–G–K–M) originates in their near-surface regions, namely in the chromosphere and corona, the origin of X-ray emission in OBA stars requires case-by-case analysis since isolated OBA stars are not intrinsic X-ray emitters.

In this work, we derive the fundamental parameters of the stars in our sample, including the effective temperature $T_{\rm eff}$ and surface gravity $\log g$, based on spectral energy distribution fitting and optical spectroscopy obtained with the 1.5-m Russian–Turkish telescope RTT-150. An additional analysis of the H$\alpha$ line profiles allows us to identify possible mechanisms responsible for the observed X-ray emission, including non-stationary stellar winds, interactions in circumstellar material, and coronal emission from hidden cool companions.

We find that the X-ray emission in eight stars, with typical luminosities in the range $\log L_{\mathrm{X}} = 28.5$–$30.0$, is most likely associated with hidden late-type companions.

\end{abstract}

\section{INTRODUCTION}

Massive stars play a key role in the evolution of galaxies, acting as major sources of ionizing radiation, kinetic energy, and heavy elements. One of the primary manifestations of their activity is X-ray emission, the properties of which depend on the stellar spectral type. For the hottest stars (O-type), X-ray emission is mainly associated with shock waves in powerful radiatively driven winds, as well as with wind interactions in close binary systems. In B-type stars, the contribution from stellar winds becomes considerably weaker; instead, magnetic fields leading to the formation of magnetospheres, together with emission from hidden companions with active coronae, may dominate. A-type stars exhibit little to no intrinsic X-ray emission due to their weak winds and the absence of convective envelopes. Consequently, any detected emission from such objects is generally attributed to hidden late-type companions \citep{Schröder2007} or, in some cases, to magnetic peculiarities.

X-ray emission in magnetic massive stars is commonly interpreted as arising from shock formation within magnetospheres, caused by collisions of stellar wind streams confined by magnetic fields. In these objects, the structure and dynamics of the stellar wind are largely governed by the magnetic field geometry. Flows emerging near the magnetic poles escape freely along open field lines into the interstellar medium, whereas at lower magnetic latitudes, charged particles are constrained to follow closed magnetic loops. As oppositely directed streams converge near the magnetic equator, they collide and form shock fronts. These regions produce hot plasma responsible for ultraviolet and X-ray emission in the surrounding region of the star \citep{Babel1997a, Babel1997b}.

The launch of the \textit{Spektr-RG} X-ray observatory \citep{Sunyaev2021} with the \textit{eROSITA} telescope \citep{Predehl2020} on board, operating in the 0.3--10\,keV range, has significantly expanded the capabilities for studying the X-ray properties of massive stars, providing wide sky coverage and high sensitivity. At the same time, reliable interpretation of the X-ray data requires a comparison with the fundamental parameters of the stars themselves, which can be refined using modern photometric, astrometric, and spectroscopic measurements.

Photometric data for the selected objects were taken from the APASS DR9 \citep{Henden2014}, ASCC \citep{Kharchenko2001}, ALL-WISE \citep{Wright2010}, 2MASS \citep{Skrutskie2006}, SDSS DR12 \citep{Alam2015}, Pan-STARRS1 \citep{Chambers2016}, GALEX \citep{Bianchi2011}, Str\"omgren Photometric Catalog \citep{Paunzen2015}, GLIMPSE \citep{Benjamin2003}, Tycho-2 \citep{Høg2000}, and \textit{Gaia} DR3 \citep{Gaia Collaboration2023} catalogs, which provide broad coverage from the ultraviolet to the infrared. Astrometric parameters (parallaxes and proper motions) were taken from the \textit{Gaia} mission, allowing us to refine the distances and select the nearest stars with the smallest uncertainties.

Spectroscopic data were obtained with the 1.5-m Russian-Turkish telescope RTT-150. The fundamental parameters of the stars in our sample were derived by fitting the observational data to model spectra and the spectral energy distribution (SED). The fitting process accounted for both the best match of the photometric data and SED to synthetic models and the detailed matching of diagnostic line profiles. For stars of spectral types B and A, the parameters were determined from the profiles of the hydrogen Balmer series lines; surface gravity was determined from the wings of these lines, which are most sensitive to this parameter. For O-type stars, the temperature was determined from the line ratios of helium in different ionization states, primarily \ion{He}{I} $\lambda$4471 and \ion{He}{II} $\lambda$4541, which serve as reliable indicators of the ionization balance in the atmosphere. The broadening of spectral lines due to rotation was accounted for using the projected rotational velocity $V\sin i$ values taken from the catalog of \citep{Glebocki2005}.

Particular attention was paid to the H$\alpha$ line profile, which allows one to distinguish between the contribution of intrinsic processes in the stellar atmosphere, manifesting as emission components, and the presence of a possible hidden companion. Such companions have a well-developed convective envelope in which a dynamo mechanism operates, leading to intense magnetic activity. The heating of coronal plasma to temperatures of about $10^6$--$10^7$\,K as a result of magnetic reconnection and flaring processes produces their X-ray emission, characteristic of young stars of late spectral types (F--M) \citep{Wright2011}. Typical X-ray luminosities for such objects are $\log L_{\mathrm{X}} \sim 28.5$--$30.0$ (in erg~s$^{-1}$) \citep{Schröder2007, Güdel2004}. According to \citep{Schröder2007}, if an A-type star is in a binary system with a late-type companion, the latter is generally a young, rapidly rotating star, leading to high magnetic activity and, accordingly, a significant level of X-ray emission. The rapid evolution of A-type stars compared to less massive late-type stars means that such companions remain relatively young and active at a stage when their more massive components have already evolved significantly.

\section{THE SELECTION OF OBA STARS}

The stellar sample was selected based on the criteria of belonging to hot stars in the upper part of the main sequence located within 3~kpc from the Sun:

\begin{enumerate}
\item Gaia absolute magnitude $M_G < 0.5$;
\item spectral classes A, B, O — i.e., stars without convective envelopes, chromospheres, and coronae, but exhibiting X-ray flux in the range $\log L_\mathrm{X} = 28.5 - 31.5$. Known accreting close binary systems with compact companions (e.g., X Per) were excluded from the sample.
\end{enumerate}

For the first group of stars targeted for spectroscopic observations with the RTT-150 telescope, 15 stars were selected; their initial parameters are given in Table~\ref{tab1}.

\begin{table*}[t]
\begin{center}
  \caption{Initial parameters of the stars}
  \label{tab1}
  \vskip 5mm
  \renewcommand{\arraystretch}{1.2}
  \renewcommand{\tabcolsep}{3pt}
  \centering
  \footnotesize
  \begin{tabular}{cccccccccccccc}
    \hline
     $N$ & HD & RA & Dec & Gmag & Sp  & $\pi$, mas & poe & $E(B-V)$ & $(B-V)o$ & $M_{Go}$ & Fx & Fxerr & Fmax/Fmin \\
     \hline
    \hline
    01 & 79158 & 09 13 48.21 & +43 13 04.2 & 5.24 & B9IV & 5.27 & 33 & 0.00 & -0.14 & -1.13 & 5.07 & 1.20 & --- \\
    02 & 112014 & 12 49 06.67 & +83 25 04.2 & 5.83 & A1IV & 8.03 & 200 & 0.00 & 0.03 & 0.37 & --- & --- & --- \\
    03 & 112028 & 12 49 13.73 & +83 24 46.4 & 5.32 & A1IVsh & 7.77 & 67 & 0.00 & -0.03 & -0.22 & 9.85 & 1.02 & 2.03 \\
    04 & 118524 & 13 34 56.53 & +70 07 08.4 & 7.50 & A0e &  2.94 & 22 & 0.00 & 0.04 & -0.14 & 1.41 & 0.49 & 1.78 \\
    05 & 137569 & 15 26 20.81 & +14 41 36.3 & 7.90 & B9Ia & 0.75 & 09 & 0.01 & -0.05 & -2.68 & 23.7 & 1.74 & 4.60 \\
    06 & 141458 & 15 48 50.48 & +12 43 25.0 & 6.80 & A0e & 4.36 & 79 & 0.05 & -0.02 & -0.17 & 2.07 & 0.62 & 2.08 \\
    07 & 157087 & 17 20 09.83 & +25 32 15.4 & 5.35 & A3III & 7.22 & 48 & 0.00 & 0.05 & -0.35 & 2.80 & 0.65 & 3.21 \\
    08 & 157857 & 17 26 17.33 & -10 59 34.8 & 7.71 & O6eV & 0.41 & 14 & 0.42 & -0.31 & -5.43 & 6.34 & 1.23 & 3.18 \\
    09 & 161693 & 17 43 59.16 & +53 48 06.1 & 5.74 & A2V & 7.03 & 87 & 0.00 & 0.01 & -0.02 & 3.64 & --- &  1.32 \\
    10 & 161677 & 17 46 41.04 & +05 46 27.4 & 7.10 & B8e & 2.88 & 93 & 0.16 & -0.17 & -1.09 & 7.57 & 1.31 & 2.10 \\
     11 & 164445 & 17 52 19.14 & +76 00 02.4 & 7.29 & F0e & 3.11 & 162 & 0.09 & 0.25 & -0.52 & 0.85 & 0.23 & 1.53 \\
    12 & 163800 & 17 58 57.25 & -22 31 03.2 & 6.89 & O7III & 0.76 & 23 & 0.49 & -0.25 & -5.18 & 11.86 & 1.68 & 2.43 \\
    13 & 164438 & 18 01 52.28 & -19 06 22.1 & 7.36 & O9IV & 0.82 & 29 & 0.55 & -0.29 & -4.78 & 9.11 & 1.47 & 5.18 \\
    14 & 174240 & 18 49 37.19 & +00 50 10.3 & 6.22 & A1IV & 5.78 & 131 & 0.00 & 0.05 & 0.05 & 7.42 & 1.35 & 2.17 \\
    15 & 182422 & 19 23 46.92 & +20 15 51.7 & 6.38 & B9V & 2.90 & 79 & 0.04 & -0.02 & -1.41 & 3.79 & 0.96 & --- \\
    \noalign{\vskip 3pt\hrule\vskip 5pt}
    \label{tab:slits}
    \end{tabular}
    \end{center}
    \vspace{-0.6cm}
 {\footnotesize   
{\bf Notes.} $\pi$, mas -- parallax in milliarcseconds from Gaia DR3; 

poe, i.e., the parallax-over-error, is the ratio of the Gaia DR3 parallax to its
measurement error;

$E(B-V)$ -- interstellar reddening. Derived from the 3D extinction maps of \cite{Green2019};

$(B-V)_0$ -- dereddened $(B-V)$ color index;

$M_{G_0}$ -- dereddened absolute magnitude in the Gaia $G$-band.
$M_{G_0} = G - 5 \log d + 5 - 1.25 \, A_G$.
The extinction $A_G$ is obtained from $A_V$ using the conversion factor of 0.843 based on the extinction curves from \citet{Zhang2023}, and an additional factor of 1.25 accounting for the temperature dependence of the $A_V \to A_G$ conversion \citep{Fouesneau2023}; the distance $d$ from \citet{Bailer-Jones2021} was used for the calculations.

Fx(--14) -- the mean X-ray flux in units of $10^{-14}~\text{erg s}^{-1}\text{cm}^{-2}$ from the eROSITA telescope measurements in the 0.3--2.3 keV energy band;

Fmax/Fmin --  the maximum-to-minimum flux ratio from the measurements in several eROSITA surveys (see Table~\ref{tab2}).} \\
\end{table*}

The results of the X-ray flux measurements from the SRG/eROSITA surveys are given in Table~\ref{tab2}.

\begin{table*}
\begin{center}
  \caption{ X-ray fluxes from the stars}
  \label{tab2}
  \vskip 5mm
  \renewcommand{\arraystretch}{1.2}
  \renewcommand{\tabcolsep}{4pt}
  \centering
  \footnotesize
  \begin{tabular}{ccccc  cc  cc  cc  cc  c}
    \hline
     $N$ & HD & SRGe & e1 & e1err & e2  & e2err & e3 & e3err & e4 & e4err & e5 & e5err & sens \\
    \hline
    \hline
    01 & 79158 & J091348.3+431307 & --- & --- & --- & --- & --- & --- & --- & --- & --- & --- & 1.13 \\
    02 & 112014 & --- &--- & --- & --- & --- & --- & --- & --- & --- & --- & --- & --- \\
    03 & 112028 & J124913.2+832448 & 6.35 & 1.75 & 12.9 & 2.8 & 10.59 & 1.95 & 9.92 & 1.89 & --- & --- & 0.58  \\
    04 & 118524 & J133454.6+700712 & --- & --- & --- & --- & 2.91 & 1.33 & --- & --- & --- & --- & 0.71  \\
    05 & 137569 & J152620.9+144136 & 44.09 & 4.71 & 9.59 & 2.47 & 11.48 & 2.94 & 34.75 & 4.88 & 14.36 & 3.38 & 0.91  \\
    06 & 141458 & J154850.6+124323 & 3.17 & 1.49 & --- & --- & --- & --- & --- & --- & 4.97 & 1.94 & 0.95  \\
    07 & 157087 & J172009.6+253215 & --- & --- & --- & --- & 5.46 & 1.88 & 3.11 & 1.26 & --- & --- & 0.71  \\
    08 & 157857 & J172617.3-105937 & 10.28 & 2.98 & 3.24 & 1.61 & --- & --- & 4.92 & 3.11 & --- & --- & 1.10  \\
    09 & 161693 & J174359.2+534804 & 3.49 & 0.90 & 4.11 & 0.87 & 3.29 & 0.98 & 3.11 & 0.78 & --- & --- & 0.4 \\
    10 & 161677 & J174641.1+054630 & 10.94 & 3.23 & 8.21 & 2.53 & 9.52 & 3.18 & 5.21 & 2.15 & --- & --- & 1.38  \\
    11 & 164445 & J175218.9+760009 & --- & --- & --- & --- & 1.05 & 0.46 & --- & --- & --- & --- & 0.34  \\
    12 & 163800 & J175857.4-223103 & 17.01 & 3.98 & 12.31 & 3.93 & 7.01 & 2.71 & 11.53 & 3.03 & --- & --- & 1.18  \\
    13 & 164438 & J180152.5-190622 & 10.47 & 3.10 & 13.69 & 3.94 & 8.32 & 2.86 & --- & --- & --- & --- &  1.16  \\
    14 & 174240 & J184937.1+005011 & --- & --- & --- & --- & 6.15 & 2.35 & 5.90 & 2.25 & --- & --- & 0.96  \\
    15 & 182422 & J192346.8+201552 & --- & --- & --- & --- & --- & --- & --- & --- & --- & --- & 0.88  \\
    \noalign{\vskip 3pt\hrule\vskip 5pt}
    \label{tab:slits}
    \end{tabular}
    \end{center}
    \vspace{-0.6cm}
{\footnotesize
{\bf Notes.} SRGe -- the source number from the SRG/eROSITA catalog;

e1, e2, e3, e4, e5 -- the fluxes in units of $10^{-14}~\text{erg s}^{-1}\text{cm}^{-2}$ registered in surveys 1--5;

e1err, e2err, e3err, e4err, e5err -- the flux errors in units of $10^{-14}~\text{erg s}^{-1}\text{cm}^{-2}$ in the corresponding surveys;

sens -- the sensitivity in units of $10^{-14}~\text{erg s}^{-1}\text{cm}^{-2}$ in the total accumulation for the sky region near the corresponding X-ray source.} \\

\end{table*}

\section{OBSERVATIONS}

The spectroscopic observations of the sample stars were carried out with the 1.5-m Russian--Turkish telescope (TUG, Antalya, Turkey) on 20--23 June 2025 using the TFOSC instrument in echelle mode. The spectra cover the wavelength range 3800--8800~\AA\ at a medium spectral resolution of about 2.5~\AA. Observations were performed with an Andor iKon-L 936 BEX2-DD-9ZQ CCD ($2048 \times 2048$ pixels, 13.5~$\mu$m pixel size) cooled to $-70^\circ$C, without binning. Two echelle exposures were obtained for each target, with exposure times between 600 and 1800~s depending on the stellar brightness. The spectra were reduced in the IRAF environment \citep{Tody1986, Tody1993}, including extraction of one-dimensional spectra, removal of cosmic rays and defects, continuum normalization, and wavelength calibration.

\section{METHOD DESCRIPTION}

To determine the fundamental parameters of the stars, we used the astroARIADNE code (spectrAl eneRgy dIstribution bAyesian moDel averagiNg fittEr; \citep{Vines2022}) based on fitting the spectral energy distribution (SED) and Bayesian model averaging over several stellar atmosphere models with weights proportional to their posterior likelihood. This approach accounts for the uncertainty associated with choosing a specific theoretical model and thus reduces systematic errors inherent in using a single grid of stellar atmosphere models.

The input data for the algorithm consist of broadband photometry corrected for interstellar extinction $A_V$ derived from interstellar dust maps (Bayestar \cite{Green2019}; dust maps \cite{Edenhofer2024} -- added by us) and the Bailer-Jones distances \citep{Bailer-Jones2021}. The code interpolates theoretical grids of stellar atmosphere models over the parameters $T_{\rm eff}$, $\log g$, and [Fe/H], computing the likelihood function by comparing observed and synthetic fluxes. Parameter estimation is performed via Markov chain Monte Carlo (MCMC) sampling, which efficiently explores the parameter space. The resulting posterior distributions provide the most probable parameter values and their confidence intervals. The stellar radius $R_\star$ is obtained by scaling the model SED to the observed photometry using the distance; the bolometric luminosity $L_{\rm bol}$ follows from the Stefan–Boltzmann law; and the stellar mass $M_\star$ is derived by interpolating the MIST evolutionary isochrones \citep{Choi2016}.

The astroARIADNE package uses the following theoretical stellar atmosphere grids: PHOENIX v2 \citep{Husser2013}, BT-Settl \citep{Allard2012}, BT-NextGen \citep{Hauschildt1999,Allard2012}, BT-Cond \citep{Allard2012}, Castelli \& Kurucz \citep{Castelli-Kurucz2003}, Kurucz \citep{Kurucz1993}, and Coelho \citep{Coelho2014}.

To adapt the code for the study of massive stars, we additionally incorporated specialised atmosphere model grids designed for hot massive stars ($T_{\rm eff} > 15\,000$ K): PoWR \citep{Pauli2025,Hainich2019}, TLUSTY \citep{Werner1999,Rauch2003,Werner2003}, and TMAP \citep{Hubeny1995,Lanz2003,Lanz2007}.

\begin{figure*}[!t]
\center{\includegraphics[width=1.0\linewidth]{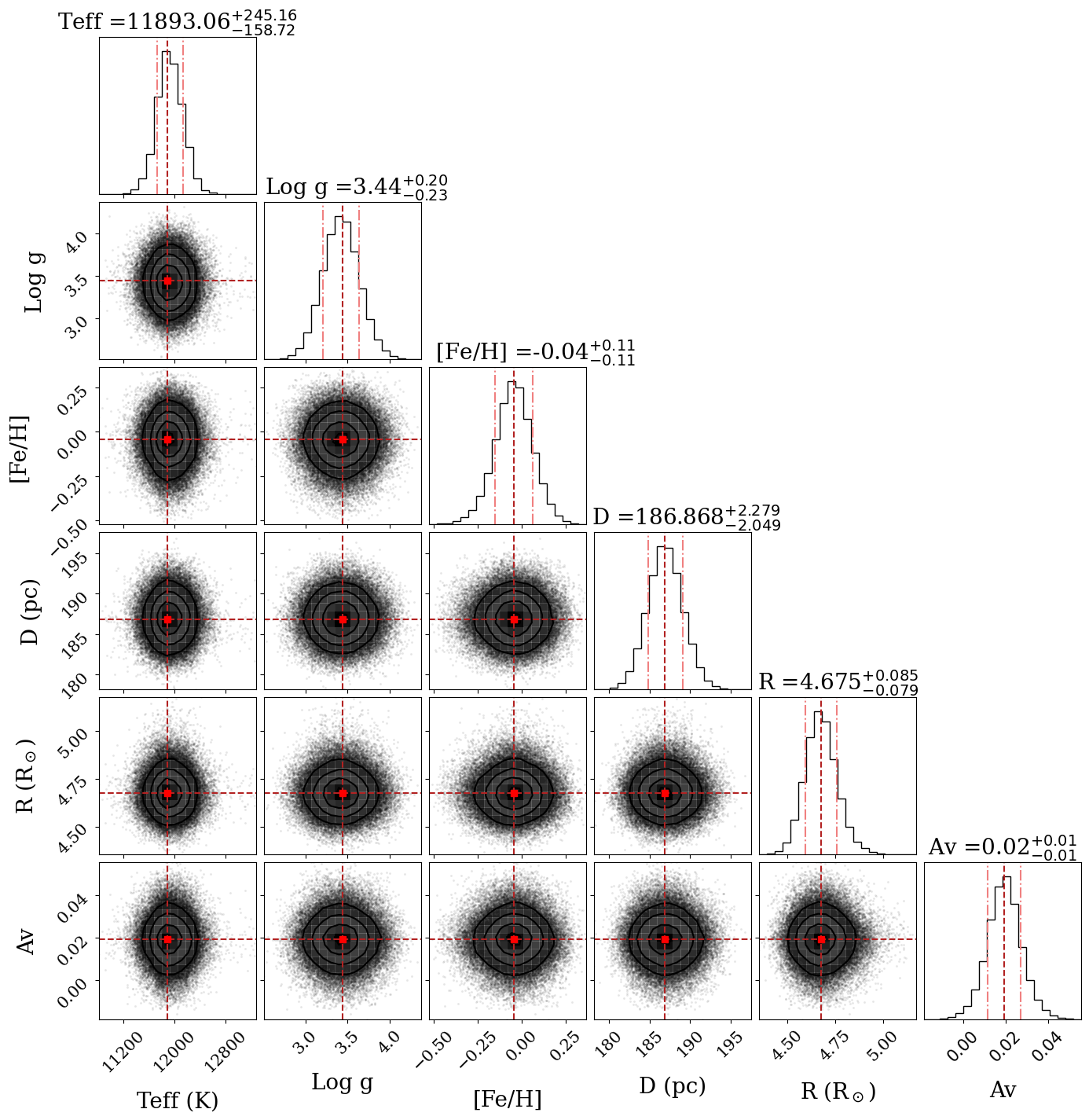}}
\caption{Corner plot showing the posterior distributions of the physical parameters for the star 36 Lyn.}
\label{corner}
\end{figure*}

\begin{figure}
\center{\includegraphics[width=1.0\linewidth]{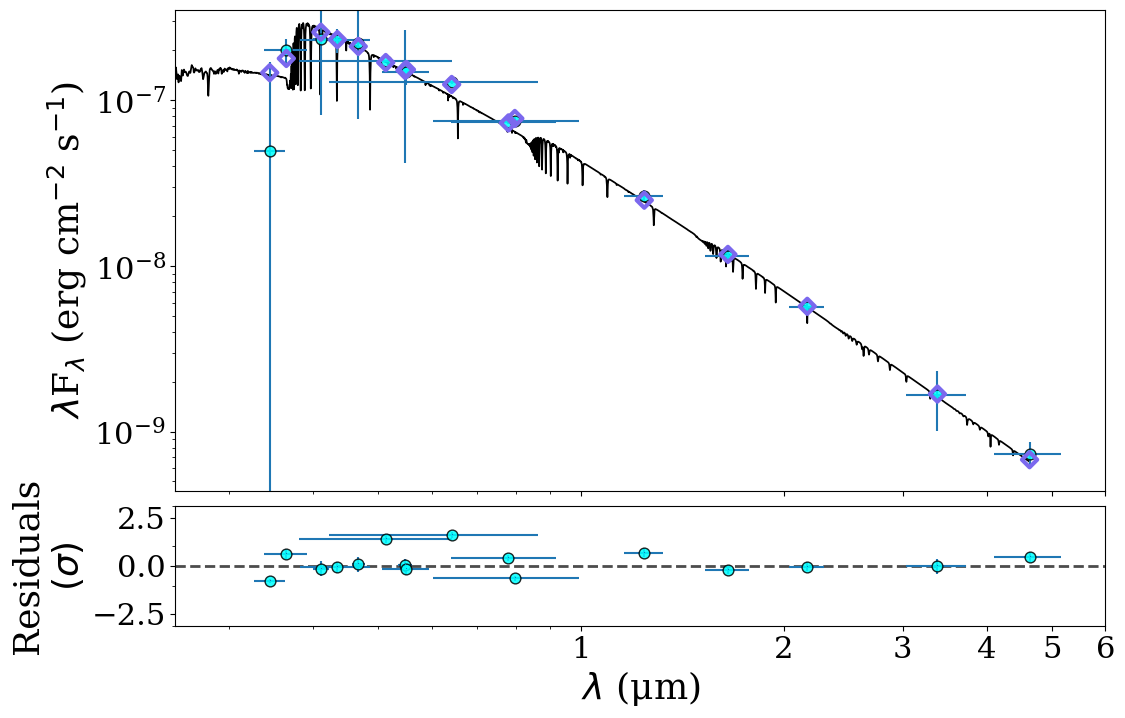}}
\caption{Top panel: the spectral energy distribution (SED) of the star 36~Lyn based on the BT-Settl model. The cyan circles correspond to the photometric observations, while the horizontal bars indicate the filter bandwidths. The violet diamonds denote synthetic photometry computed from the model spectrum.
Bottom panel: The residuals of the model from the photometric observations, normalized by their errors.}
\label{SED}
\end{figure}

\begin{figure}[!t]
\center{\includegraphics[width=1.0\linewidth]{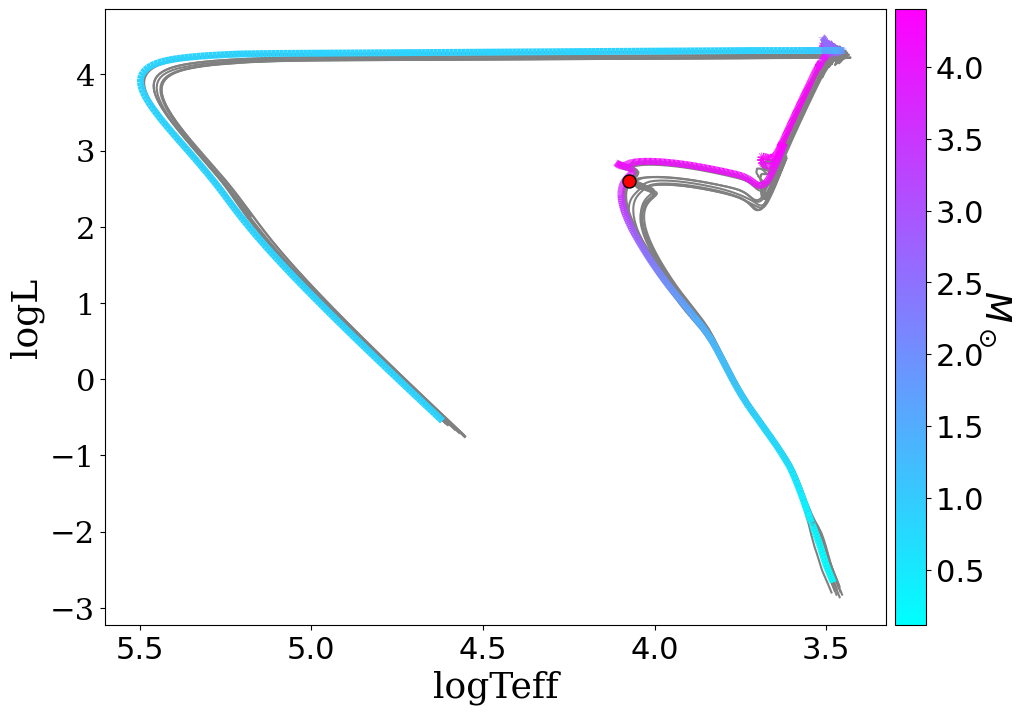}}
\caption{The Hertzsprung--Russell diagram for 36~Lyn.
Constructed from the MIST isochrone grid interpolated to the best-fit stellar parameters. The vertical and horizontal error bars are smaller than the symbol size. The gray lines correspond to randomly selected parameter realizations from the posterior distribution, reflecting the fitting uncertainties.}
\label{HR}
\end{figure}

The output of the code is illustrated for the star 36 Lyn in the following figures:

-- the corner diagram showing the posterior distributions of the physical parameters (Fig.~\ref{corner}),

-- the spectral energy distribution fitted with the most probable (highest-likelihood) theoretical atmosphere model (Fig.~\ref{SED}),

-- the position of the best-fitting parameters on the Hertzsprung–Russell diagram (Fig.~\ref{HR}), and

\begin{figure*}[h]
\begin{minipage}{0.5\linewidth}
\center{\includegraphics[width=1\linewidth]{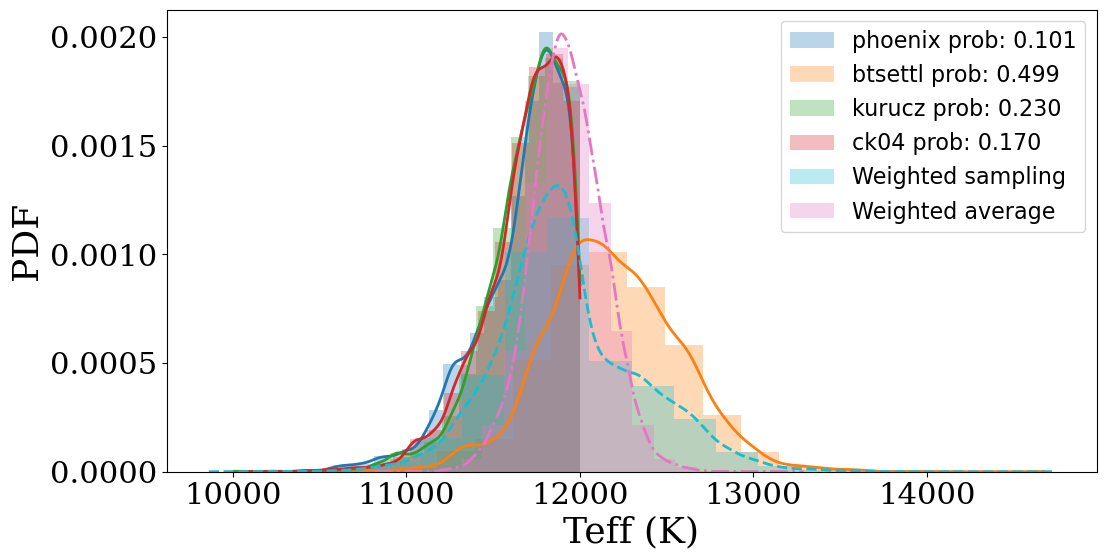}}
\end{minipage}
\hfill
\begin{minipage}{0.5\linewidth}
\center{\includegraphics[width=1\linewidth]{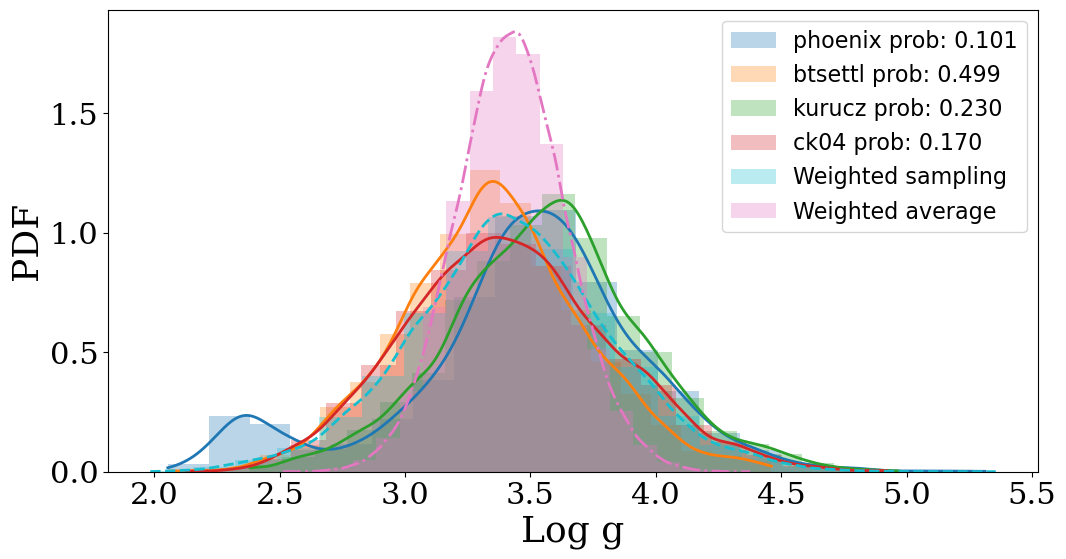}}
\end{minipage}
\caption{The posterior distributions of $T_{\rm eff}$ and $\log g$ for 36~Lyn. The samples from the posterior distributions obtained with astroARIADNE for different model grids are shown. The posterior probability is indicated for each model.}
\label{teff_logg}
\end{figure*}

-- the posterior samples of $T_{\rm eff}$ and $\log g$ with individual model probabilities, and the averaged posterior samples (Fig.~\ref{teff_logg}).

The parameters derived for all stars are listed in Table~\ref{tab3}. A detailed discussion of the obtained parameters, including an analysis of the possible presence of hidden companions and the origin of X-ray emission for individual stars, is presented in the following sections.

\section{THE NATURE OF THE X-RAY EMISSION FROM INDIVIDUAL SOURCES}

\subsection{36~Lyn (=HD~79158)}

\begin{figure*}
\center{\includegraphics[width=1.0\linewidth]{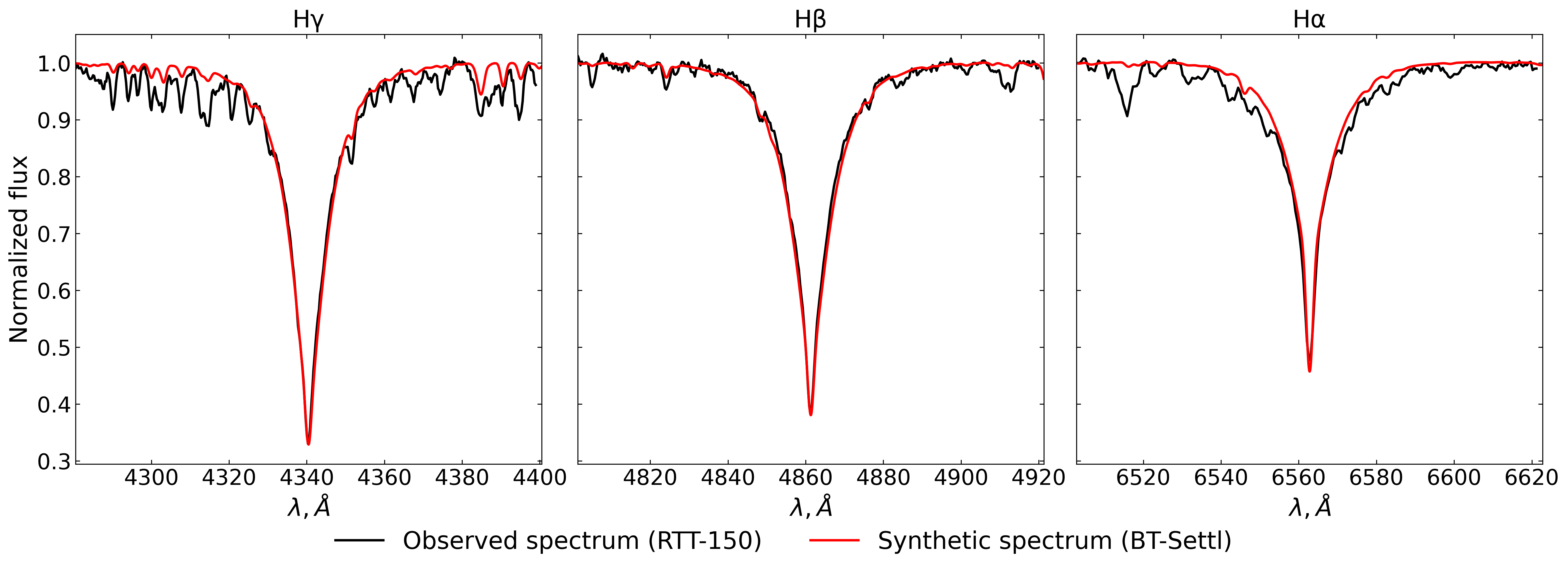}}
\caption{Comparison of the observed and theoretical spectra (based on the BT-Settl model) for the star 36~Lyn.}
\label{Hb_Ha_36Lyn}
\end{figure*}

36~Lyn is a bright ($V=5.28^m$), well-studied, chemically peculiar magnetic star of late B type (B8p) with weak helium lines. \citet{Wade2006} performed a comprehensive analysis using polarimetric and spectroscopic observations. They derived the stellar parameters ($T_{\rm eff} \approx 13\,000$--$13\,600$~K, $\log g \approx 3.7$--$4.2$, $\log L/L_\odot = 2.54 \pm 0.16$, $R_* = 3.4 \pm 0.7\,R_\odot$, $B_p = 3.57 \pm 0.36$~kG), confirmed the variability of the H$\alpha$ line core profile due to the occultation of the core by circumstellar gas, and interpreted the spread in radial velocity measurements ranging from $+21$~km~s$^{-1}$ \citep{Hoffleit1991} to $29.7 \pm 1.6$~km~s$^{-1}$ \citep{Takada-Hidai1989} as being due to the variability of the metal line profiles caused by their nonuniform distribution on the stellar surface and the variability of the H$\alpha$ line profile due to the presence of circumstellar gas. During our spectroscopic observations, the star was at a phase where the emission component in the H$\alpha$ line was absent (Fig.~\ref{Hb_Ha_36Lyn}).

According to \citet{Berghoefer1996}, the star 36~Lyn was not reliably detected by the ROSAT X-ray observatory, and only an upper limit on its X-ray luminosity was given: $\log L_\mathrm{X} < 29.84$. The SRG/eROSITA X-ray telescope detected X-ray emission from the star with $\log L_\mathrm{X} = 29.33$, which is slightly below that upper limit. There are two possible explanations for the origin of the X-ray emission in the system:

\begin{enumerate}
\item Heating of material and generation of X-rays in closed magnetic loops formed by the interaction of the magnetic field with the stellar wind;
\item The X-ray emission originates from a companion -- an F0 star discovered by speckle interferometry \citep{Balega2012}. \citet{Balega2012} estimated the orbital period of 20 years, implying a wide binary system with no interaction between the components.
\end{enumerate}

\begin{table*}[t]
\begin{center}
\footnotesize
\caption{Parameters of the stars}\label{tab3}
\setlength{\tabcolsep}{3pt}
\renewcommand{\arraystretch}{1.5}
\begin{tabular}{lrrrrrrrrrrr}
\hline
Star & \makecell{$d_{\rm BJ}$ \\ (pc)} & \makecell{$T_{\rm eff}$ \\ (K)} &
\makecell{$\log g$ \\ (dex)} & \makecell{$R_\star$ \\ ($R_\odot$)} & \makecell{Age \\ ($10^6$ yr)} & \makecell{$M_\star$ \\ ($M_\odot$)} & \makecell{$L_{\rm bol}$ \\ ($L_\odot$)} &
\makecell{$\log L_\mathrm{X}$}\textsuperscript{**} & \makecell{$\log L_\mathrm{X}$ \\ (СРГ)} & $\log\!\left(\frac{L_\mathrm{X}}{L_{\rm bol}}\right)$ & Comp. \\
\hline
\hline
36 Lyn & $\textbf{188}^{+5}_{-5}$ & $\textbf{11893}^{+380}_{-281}$ & $\textbf{3.4}^{+0.3}_{-0.4}$ & $\textbf{4.7}^{+0.1}_{-0.1}$ & $\textbf{161}^{+86}_{-25}$ & $\textbf{3.5}^{+0.6}_{-0.1}$ & $\textbf{396}^{+54}_{-45}$ & < 29.84 & 29.33 & -6.85 & $\checkmark$? \\
HD 112014 & $\textbf{124.2}^{+0.6}_{-0.5}$ & $\textbf{9630}^{+203}_{-172}$ & $\textbf{3.9}^{+0.3}_{-0.3}$ & $\textbf{3.0}^{+0.1}_{-0.1}$ & $\textbf{430}^{+134}_{-100}$ & $\textbf{2.5}^{+0.2}_{-0.3}$ & $\textbf{69}^{+7}_{-6}$ & 28.63 ? &  &  &  \\
HD 112028 & $\textbf{128}^{+2}_{-2}$ & $\textbf{9983}^{+81}_{-89}$ & $\textbf{3.6}^{+0.2}_{-0.2}$ & $\textbf{5.5}^{+0.1}_{-0.1}$ & $\textbf{240}^{+37}_{-25}$ & $\textbf{3.6}^{+0.2}_{-0.2}$ & $\textbf{275}^{+16}_{-18}$ & 29.01 & 29.29 &  & $\checkmark$ \\
HD 118524 & $\textbf{338}^{+15}_{-13}$ & $\textbf{8839}^{+168}_{-165}$ & $\textbf{3.5}^{+0.3}_{-0.3}$ & $\textbf{2.3}^{+0.1}_{-0.1}$ & $\textbf{731}^{+480}_{-129}$ & $\textbf{2.0}^{+0.1}_{-0.4}$ & $\textbf{30}^{+3}_{-3}$ &  & 29.29 &  & $\checkmark$ \\
HD 137569 & $\textbf{1288}^{+140}_{-134}$ & $\textbf{11989}$\textsuperscript{***} & $\textbf{2.00}$\textsuperscript{***} & $\textbf{43}^{+5}_{-15}$ & $\textbf{88}^{+8}_{-8}$ & $\textbf{5.3}^{+0.2}_{-0.2}$ & $\textbf{33988}^{+7314}_{-19748}$ & 31.21 & 31.67 & -6.44 &  \\
HD 141458 & $\textbf{228}^{+3}_{-2}$ & $\textbf{9036}^{+283}_{-192}$ & $\textbf{3.4}^{+0.3}_{-0.3}$ & $\textbf{3.9}^{+0.1}_{-0.1}$ & $\textbf{574}^{+44}_{-57}$ & $\textbf{2.3}^{+0.1}_{-0.1}$ & $\textbf{92}^{+12}_{-10}$ &  & 29.11 &  & $\checkmark$ \\
HD 157087 & $\textbf{138}^{+3}_{-3}$ & $\textbf{8488}^{+178}_{-136}$ & $\textbf{3.4}^{+0.2}_{-0.2}$ & $\textbf{4.8}^{+0.1}_{-0.1}$ & $\textbf{419}^{+118}_{-36}$ & $\textbf{2.9}^{+0.1}_{-0.3}$ & $\textbf{110}^{+11}_{-10}$ &  & 28.81 &  & $\checkmark$ \\
HD 161677 & $\textbf{341}^{+3}_{-3}$ & $\textbf{14804}^{+489}_{-1206}$ & $\textbf{3.7}^{+0.2}_{-0.3}$ & $\textbf{3.8}^{+0.2}_{-0.1}$ & $\textbf{122}^{+33}_{-43}$ & $\textbf{4.2}^{+0.5}_{-0.3}$ & $\textbf{619}^{+131}_{-177}$ &  & 30.02 & -6.35 &  \\
HD 161693 & $\textbf{142}^{+1}_{-2}$ & $\textbf{9043}^{+199}_{-162}$ & $\textbf{3.4}^{+0.2}_{-0.2}$ & $\textbf{3.9}^{+0.1}_{-0.1}$ & $\textbf{580}^{+40}_{-135}$ & $\textbf{2.3}^{+0.3}_{-0.1}$ & $\textbf{92}^{+10}_{-8}$ & 29.32 ? & 28.94 &  & $\checkmark$ \\
HD 164445 & $\textbf{319}^{+2}_{-2}$ & $\textbf{6909}^{+67}_{-71}$ & $\textbf{3.4}^{+0.4}_{-0.4}$ & $\textbf{6.7}^{+0.2}_{-0.1}$ & $\textbf{549}^{+25}_{-36}$ & $\textbf{2.5}^{+0.1}_{-0.1}$ & $\textbf{93}^{+6}_{-5}$ &  & 29.01 & -6.54 &  \\
HD 174240 & $\textbf{172}^{+1}_{-1}$ & $\textbf{8942}^{+192}_{-214}$ & $\textbf{3.6}^{+0.1}_{-0.1}$ & $\textbf{3.9}^{+0.1}_{-0.1}$ & $\textbf{456}^{+80}_{-48}$ & $\textbf{2.6}^{+0.2}_{-0.3}$ & $\textbf{87}^{+9}_{-8}$ & 29.34 & 29.42 &  & $\checkmark$ \\
HD 182422 & $\textbf{340}^{+4}_{-5}$ & $\textbf{9362}^{+360}_{-311}$ & $\textbf{2.9}^{+0.2}_{-0.2}$ & $\textbf{7.1}^{+0.2}_{-0.2}$ & $\textbf{234}^{+31}_{-25}$ & $\textbf{3.5}^{+0.2}_{-0.2}$ & $\textbf{346}^{+60}_{-47}$ & < 29.55 & 29.72 &  & $\checkmark$ \\
HD 157857 & $\textbf{2272}^{+145}_{-163}$ & $\textbf{36094}^{+916}_{-916}$ & $\textbf{3.8}^{+0.3}_{-0.5}$ & $\textbf{13.4}^{+1.3}_{-1.1}$ & $\textbf{4}^{+1}_{-1}$ & $\textbf{27.4}^{+5.4}_{-2.9}$ & $\textbf{270692}^{+66973}_{-47981}$ & < 32.76\textsuperscript{*} & 31.59 & -7.42 &  \\
HD 163800 & $\textbf{1250}^{+55}_{-45}$ & $\textbf{34629}^{+1403}_{-1317}$ & $\textbf{3.7}^{+0.1}_{-0.1}$ & $\textbf{11.1}^{+0.6}_{-0.6}$ & $\textbf{4}^{+3}_{-2}$ & $\textbf{23.0}^{+4.5}_{-3.8}$ & $\textbf{159374}^{+33043}_{-29155}$ &  & 31.35 & -7.44 &  \\
HD 164438 & $\textbf{1166}^{+32}_{-30}$ & $\textbf{31977}^{+580}_{-774}$ & $\textbf{3.8}^{+0.1}_{-0.2}$ & $\textbf{10.2}^{+0.7}_{-0.6}$ & $\textbf{4}^{+2}_{-3}$ & $\textbf{18.3}^{+1.8}_{-1.5}$ & $\textbf{96747}^{+16356}_{-14869}$ &  & 31.17 & -7.40 &  \\
\hline
\end{tabular}
\end{center}
{\footnotesize
{\bf Notes.} $d_{\rm BJ}$ -- the Bailer-Jones distance \citep{Bailer-Jones2021}.

\textsuperscript{*} The logarithm of X-ray luminosity obtained from the Einstein observatory data \citep{Chlebowski1989}. 

\textsuperscript{**} The logarithm of X-ray luminosity obtained from the ROSAT observatory data \citep{Schröder2007}. The question mark denotes the stars from the list of A-type stars in binary or multiple systems associated with the X-ray emission for which there is evidence of hidden companions. 

\textsuperscript{***} The value was derived from the spectroscopic data and fixed when fitting the SED fitting, because the spectral energy distribution did not allow reliable determination of the parameter due to a significant contribution from infrared radiation caused by the presence of a disk in the system.

In the last column, the symbol ``$\checkmark$'' indicates the presence of a hidden companion in the system.}
\end{table*}

\begin{figure*}[!t]
\center{\includegraphics[width=1.0\linewidth]{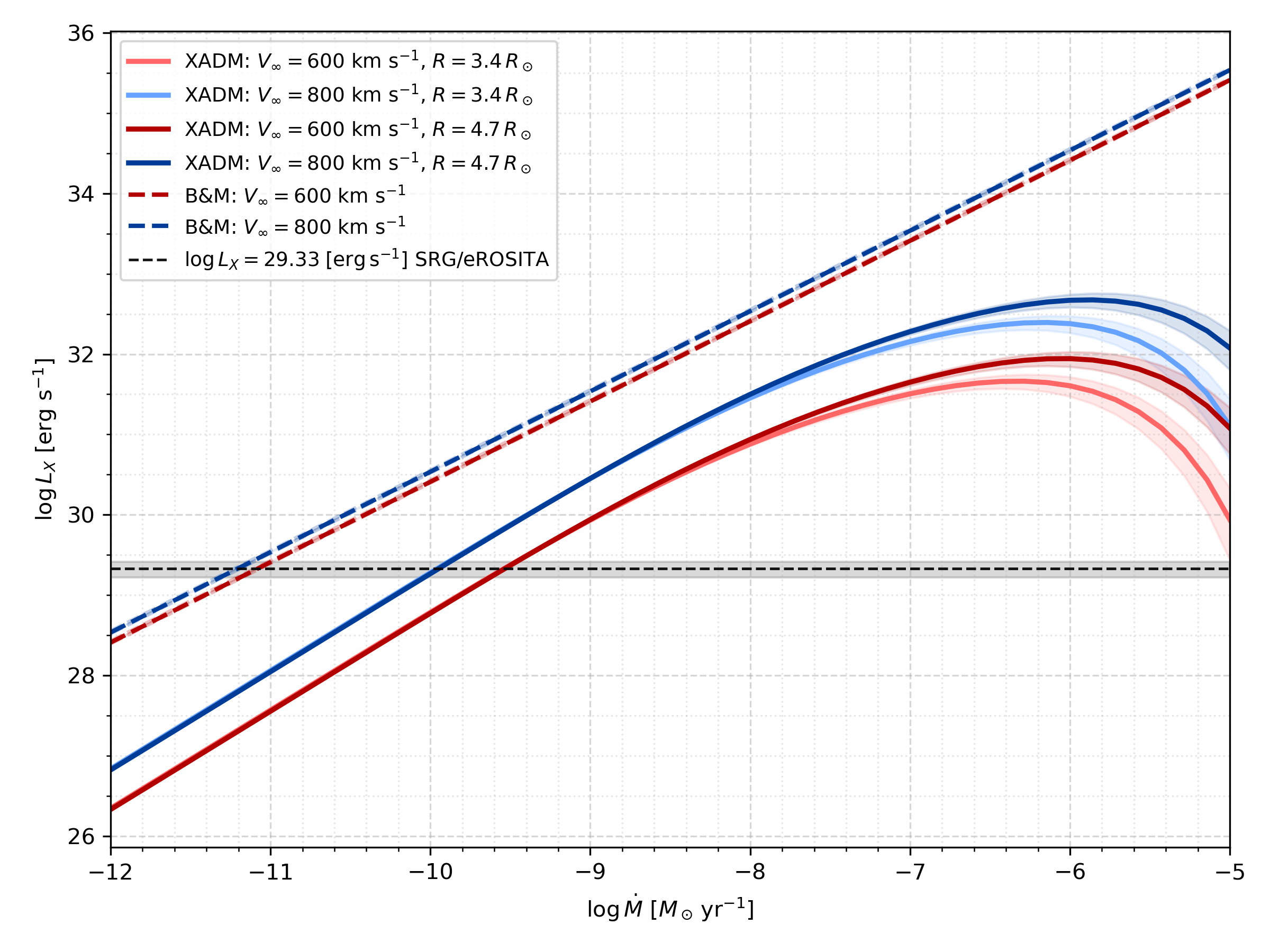}}
\caption{Calculated X-ray luminosity as a function of mass-loss rate using the analytic formula of the XADM model (X-ray Analytic Dynamical Magnetosphere, \citep{ud-Doula2014}) -- solid line, and the power-law relation of Babel \& Montmerle (1997a) -- dashed line for several sets of parameters $V_\infty$ and $R_*$. The magnetic field strength is taken to be $B=3.57 \pm 0.36$ kG \citep{Wade2006}. The radius value $R_* = 3.4 R_\odot$ is taken from \citep{Wade2006}, while $R_* = 4.7 R_\odot$ is obtained in the present work. As can be seen from the plot, the model curves derived for different radii begin to diverge noticeably only at large values of $\dot{M}$, whereas in the region of the observed X-ray luminosity these differences are practically absent. The transparent area illustrates the uncertainty range.}
\label{Lx_Mdot}
\end{figure*}

The X-ray luminosity of magnetic stars depends on several parameters: the magnetic field strength $B$, the mass-loss rate $\dot{M}$, the terminal wind velocity $V_\infty$, and the stellar radius $R_\star$. Figure~\ref{Lx_Mdot} shows the calculated X-ray luminosity of 36~Lyn as a function of mass-loss rate using the analytical formula of the XADM model (X-ray Analytic Dynamical Magnetosphere; \citealp{ud-Doula2014}). This model accounts for wind velocity, magnetic confinement, shock cooling, and flow geometry. For comparison, the dashed line shows the power-law relation from Babel \& Montmerle (1997a), given by the formula:

\begin{equation*}
L_\mathrm{X} = 2.6 \times 10^{30}~\text{erg~s}^{-1}~\dot{M}_{-10}~V_8~B_3^{0.4},
\end{equation*}

\begin{align*}
\dot{M}_{-10} &\equiv \frac{\dot{M}}{10^{-10}\,M_\odot\,\text{yr}^{-1}}, \\
V_8 &\equiv \frac{V_\infty}{10^8~\text{sm\,s}^{-1}}, \\
B_3 &\equiv \frac{B_p}{10^3~\text{G}}.
\end{align*}

Calculations using the ud-Doula formula show that for sufficiently high mass-loss rates and terminal velocities, X-ray luminosities comparable to those observed for 36~Lyn can be achieved.

The second scenario is also plausible, since young F0 stars can have X-ray luminosities as high as $L_\mathrm{X} \lesssim 10^{30}~\text{erg s}^{-1}$. The scatter in radial velocity measurements is most likely related to a hidden companion.

Thus, both scenarios remain admissible interpretations of the observed X-ray luminosity.

\vspace{1.5em}
\subsection{HD~112028 (=HR~4893) and HD~112014 (=HR~4892)}

Although HD~112028 and HD~112014 have traditionally been regarded as a multiple system, their large projected separation ($\approx 4$~pc) renders a physical association unlikely. HD~112028 is classified as a rapidly rotating A1~III shell star whose envelope results from fast rotation. Stars of this class typically exhibit two sets of spectral lines: broad features formed in the hot photosphere and narrow lines (Ca~II, H~I, Fe~II) arising in a cooler, extended envelope (see, e.g., \citealp{Andrillat1986}). 

Like other shell-type A stars, HD~112028 displays an infrared excess due to emission from cool circumstellar gas or dust. In the spectral energy distribution (SED), this excess appears as a flux elevation above the photospheric level in the near-infrared, most notably in the 2MASS $H$ and $K$ bands. To ensure a reliable SED fit, these bands were excluded from the fitting.

HD~112014, on the other hand, is a spectroscopic binary consisting of two main-sequence stars, A0~V + A2~V, that orbit their common centre of mass with a period of $P_{\rm orb} \sim 3.3$~d on a nearly circular orbit ($e \approx 0.04$) \citep{Eggleton2008}. In our spectra, lines of only one component are present.

According to the Schr\"oder \& Schmitt catalog of A stars with hidden companions \citep{Schröder2007}, HD~112028 is included among the detected ``bona fide'' single A-type stars reliably associated with X-ray sources ($\log L_\mathrm{X} = 29.01$), strongly suggesting the presence of a hidden late-type companion. HD~112014, by contrast, appears among the X-ray associated A-type stars that are members of known binary or multiple systems or show signs of hidden companions, with an X-ray luminosity of $\log L_\mathrm{X} = 28.63$.

New measurements with the SRG/eROSITA telescope have made it possible to unambiguously establish that the X-ray source is the star HD~112028, given the positional accuracy of the X-ray coordinates of $5''$. At the position of the star HD~112014, no X-ray flux is detected at the sensitivity level of $5.83 \times 10^{-15}~\text{erg~s}^{-1}\text{cm}^{-2}$.

The X-ray luminosity of HD~112028 measured by SRG/eROSITA is $\log L_\mathrm{X} = 29.29$. Such a level of X-ray emission is typical of late-type hidden companions. Moreover, the spectrum of HD~112028 exhibits no signatures of peculiarity or a magnetic field that would explain such X-ray luminosity. Consequently, HD~112028 is at least a binary system, whereas the double star HD~112014 is not the source of the X-ray emission.

\subsection{HD~118524}
HD~118524 is an ordinary A-type star. The X-ray luminosity detected by the SRG/eROSITA telescope is $\log L_\mathrm{X} < 29.29$ and indicates the presence of a hidden companion. According to \textit{Gaia} data, the $RUWE = 4.75$ value significantly exceeds the threshold $RUWE_{\rm threshold} = 1.23$ \citep{Castro-Ginard2024}, further supports the existence of a secondary component.

\subsection{HD~137569}

HD~137569 is a binary star located at a high Galactic latitude and belongs to the old Galactic population. Orbital parameters of the system are known from the 9th Catalogue of Spectroscopic Binary Orbits \citep{Pourbaix2004}: orbital period $P_{\rm orb}=529.8$~d, eccentricity $e=0.11$, and radial-velocity semi-amplitude $K_1 = 15.8$~km\,s$^{-1}$.  The primary star of the binary system is in the post-asymptotic giant branch (post-AGB) stage and has an extremely low metallicity $[\mathrm{Fe}/\mathrm{H}] = -3$ \citep{Kluska2022}.

\begin{figure*}
\center{\includegraphics[width=1.0\linewidth]{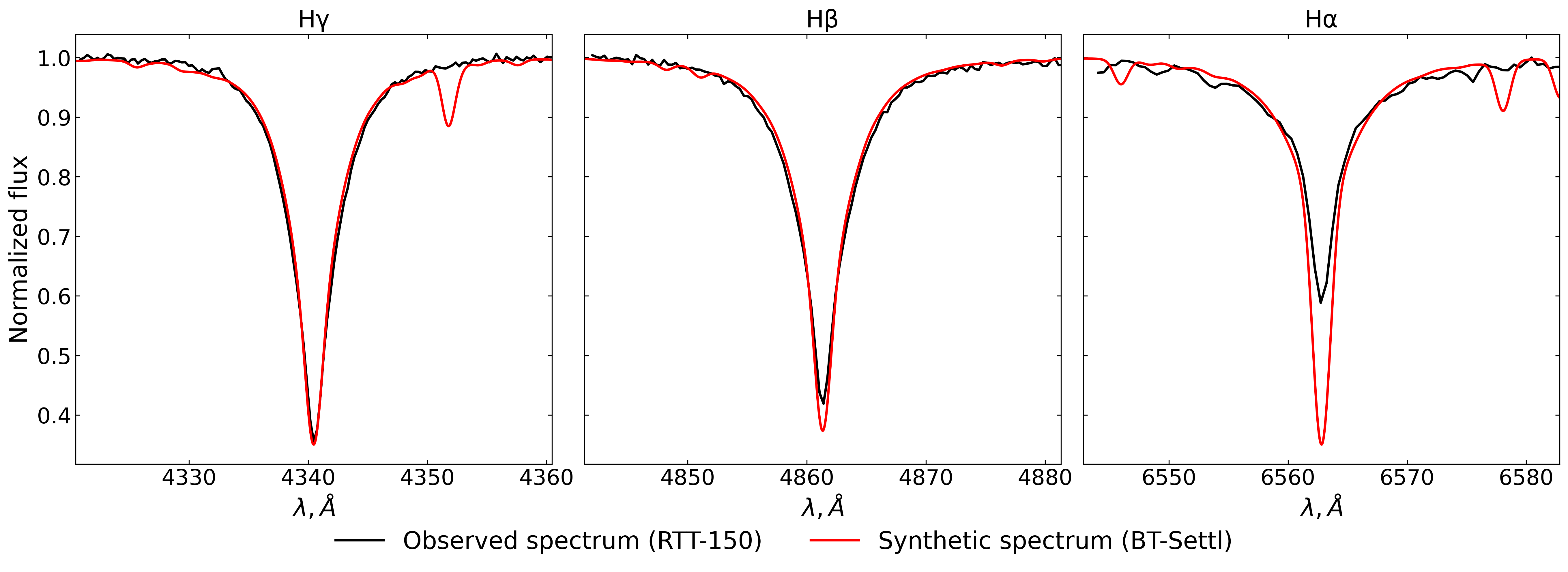}}
\caption{Comparison of the observed and synthetic spectra ((based on the BT-Settl model) for HD~137569.}
\label{Ha_Hb_Hg_HD137569}
\end{figure*}

The binary system is surrounded by a gas-dust disk \citep{Kluska2022}, whose emission manifests as a slight infrared excess and an emission contribution in the H$\alpha$ line and partially in the H$\beta$ line (Fig.~\ref{Ha_Hb_Hg_HD137569}). The ROSAT satellite detected emission with a hardness ratio $\mathrm{HR} = -1$, indicating very soft X-rays, and a flux $F_\mathrm{X} = 8.12 \times 10^{-14}~\text{erg~s}^{-1}\,\text{cm}^{-2}$ \citep{Freund2022}. The flux measured by the SRG/eROSITA telescope is $F_\mathrm{X} = 2.37 \times 10^{-13}~\text{erg~s}^{-1}\,\text{cm}^{-2}$, corresponding to an X-ray luminosity of $\log L_\mathrm{X} = 31.7$.

The pronounced underabundance of metals detected in the stellar atmosphere can be interpreted as evidence of re-accretion of gas depleted in dust from the circumbinary disk onto the post-AGB star \citep{Oomen2019}, since heavy elements preferentially condense onto dust grains in the circumbinary disk. The combination of a soft X-ray component ($\log L_\mathrm{X} > 31$, $\mathrm{HR} = -1$) and the filled H$\alpha$ profile suggests that this process likely continues at the present time, manifesting itself as low-efficiency accretion of residual gaseous disk material.

\subsection{HD~141458}

The HD~141458 system is a well-studied spectroscopic binary. Its orbital elements were first determined by Bolton et al.~(\citeyear{Bolton1983}), who derived radial velocity curves for both components and obtained a full orbital solution. The system has an orbital period of $P = 28.949$~d and a significant eccentricity of $e = 0.64$. The semi-amplitudes of the radial velocity curves are $K_1 = 60.6$~km~s$^{-1}$ and $K_2 = 66.6$~km~s$^{-1}$, corresponding to a mass ratio close to unity, $m_1/m_2 = 1.1$.

Both stars belong to spectral class~A: the primary is classified as A0V and the secondary as A1V. In our spectra, only the A1V component is reliably detected. The orbital parameters derived by Bolton et al. were later confirmed and included in the 9th Catalogue of Spectroscopic Binary Orbits (SB9; \citealp{Pourbaix2004}). \cite{Horch2017} measured a separation between the components of $0.4923''$ using speckle interferometry, with a magnitude difference of $3.5$~mag.

The X-ray luminosity measured by SRG/eROSITA is $L_\mathrm{X} = 1.29 \times 10^{29}~\text{erg~s}^{-1}$. Since both visible components are A-type stars and therefore cannot be X-ray sources, the most probable explanation is a hidden late-type companion whose emission is produced by an active corona.

\subsection{HD~157087}

HD~157087 was studied in detail by \cite{Khalack2018}, who refuted its classification as an Am star \citep{Yüce2011,Preston1974,Adelman1987}. The observed features include the variability of the mean abundances of some chemical elements against an almost constant background of the others, a significant enhancement of C, S, Ca, Sc, V, Cr, Mn, Co, Ni, and Zr with depth in the atmospheric layers, and overabundances of Ca and Sc. The author interprets all of these as evidence that the star is chemically peculiar, with characteristics close to those of Am stars but exhibiting pronounced vertical stratification.

Analysis of radial velocity measurements revealed both long-period and short-period variations. 
The long-period variations support the suggestion that HD~157087 is an astrometric binary with a period exceeding 6~yr \citep{Makarov2005}. 
The presence of short-period radial velocity variations, together with temporal changes in the mean chemical abundances, indicates that HD~157087 may be a triple system in which a short-period binary orbits a third star. 
In this case, the short-period pair may consist of two slowly rotating stars of spectral types Am and A (or a weakly magnetic Ap star), with similar effective temperatures and $\log\,g$ values but different chemical peculiarities.

The SRG/eROSITA X-ray data likewise point to the multiplicity of the system. 
Its X-ray luminosity $L_\mathrm{X} = 6.39 \times 10^{28}~\text{erg~s}^{-1}$ stems from a hidden low-mass companion, as no magnetic field is detected in the system \citep{Khalack2018}. 
The fundamental parameters we derive for the system fully agree with those reported by \cite{Khalack2018} ($T_{\rm eff} = 8882$~K, $\log\,g = 3.57$) and correspond to a star of spectral type A3\,III.

\subsection{HD~161677}

\begin{figure}[h]
\center{\includegraphics[width=1.0\linewidth]{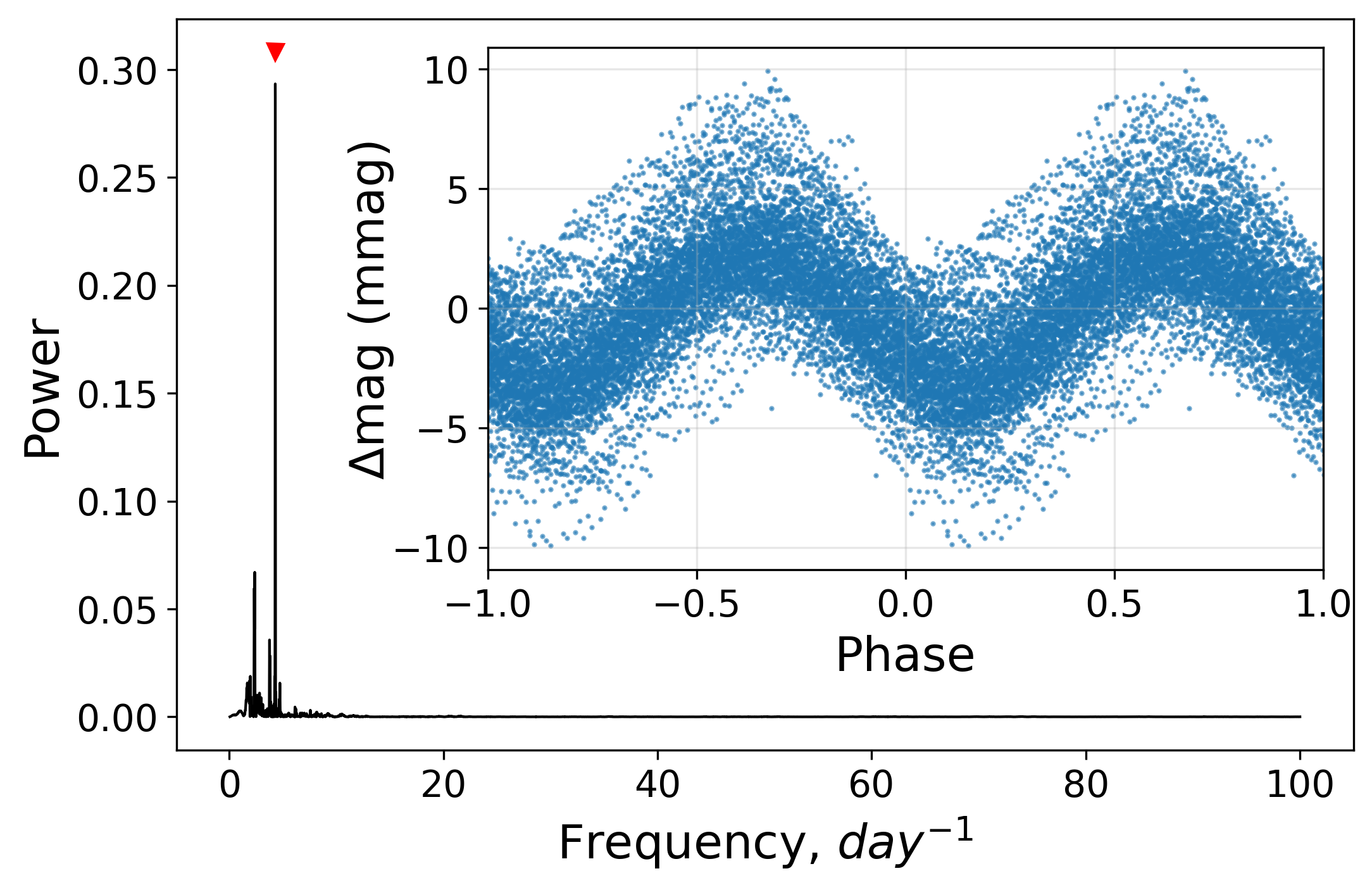}}
\caption{Periodogram and phase-folded light curve of HD~161677 based on TESS data (Sector 80). The main panel shows the Lomb--Scargle periodogram (Lomb, 1976; Scargle, 1982), with the dominant period $P_{\rm rot} = 0.233$~d. The inset displays the phase-folded light curve using this period. The fluxes are converted into relative magnitudes.}
\label{tess_lc_HD161677}
\end{figure}

HD~161677 is a rapidly rotating single B5V star with a projected rotational velocity of $V\sin i = 210$~km~s$^{-1}$ \citep{Glebocki2005} and a rotation period of $P_{\rm rot} = 0.233$~d (Fig.~\ref{tess_lc_HD161677}).
It is a member of the small, young open cluster IC~4665 \citep{Negueruela2024}.

The cluster was observed with the ROSAT satellite \citep{Giampapa1998}; however, no X-ray emission from HD~161677 was detected.
According to SRG/eROSITA observations, the X-ray luminosity of the star is $\log L_\mathrm{X} = 30.02$.

As in other rapidly rotating B-type stars in the cluster, the X-ray emission of HD~161677 is produced by shock heating in a radiatively driven stellar wind.
Inhomogeneous gas flows interact, giving rise to shocks with temperatures of several million kelvin, which generate soft X-ray emission.

\subsection{HD~161693 (=Alruba)}

HD~161693 is a normal main-sequence A1 star \citep{Zorec2012} and shows no intrinsic properties that could account for detectable X-ray emission.
Nevertheless, in the catalog of X-ray A stars in binary
and multiple systems \citep{Schröder2007}, based on ROSAT data, the star was classified as a candidate object with a hidden companion.

In this catalogue, HD~161693 is given an X-ray luminosity of $L_\mathrm{X} = 2.08 \times 10^{29}~\text{erg~s}^{-1}$, comparable to the SRG/eROSITA estimate of ($L_\mathrm{X} = 0.88 \times 10^{29}~\text{erg~s}^{-1}$).

The SRG/eROSITA observations, benefiting from higher astrometric accuracy, allow a secure identification of the X-ray source with HD~161693.
This supports the interpretation that the X-ray emission originates from a hidden low-mass companion.

\subsection{HD~164445}

We classify this object as an F2V star with a projected rotational velocity of $V\sin i = 120$~km~s$^{-1}$.
According to \textit{Gaia} astrometry, the renormalised unit weight error, $\mathrm{RUWE} = 1.121$, is below the sky-varying threshold of $\mathrm{RUWE}_{\mathrm{threshold}} = 1.229$ defined in \cite{Castro-Ginard2024}, and therefore shows no evidence of an astrometric signature of a hidden companion.

Early F-type stars possess relatively shallow convective envelopes, which limit the efficiency of the magnetic dynamo and typically result in low levels of coronal activity.
At the same time, the high rotational velocity of the star ($V\sin i = 120$~km~s$^{-1}$) implies a small Rossby number, making the dynamo more efficient and thereby enhancing both magnetic activity and X-ray emission.

According to \cite{Wright2011}, F-type stars do not exhibit a well-defined saturation regime: as the Rossby number decreases, their coronal activity passes straight from the supersaturated to the unsaturated regime without forming an extended plateau.

Given the above, the observed X-ray luminosity of this object, $L_\mathrm{X} = 1.03 \times 10^{29}~\text{erg~s}^{-1}$, is most likely driven by a combination of its relatively rapid rotation and a still-operating, though weakened, magnetic dynamo in a thin convective envelope.

\subsection{HD~174240}

The star HD~174240 has been included in several high-quality spectral libraries.
The \textit{HST Low-Resolution Stellar Library} (LOWLIB) \citep{Pal2023} provides a STIS spectrum covering $0.2{-}1.0~\mu$m, with an \ion{Mg}{II}\,$\lambda2800$ index of $0.194$~mag, indicative of photospheric absorption and a lack of pronounced chromospheric activity.
The stellar parameters derived from this library are $T_{\rm eff}=8879$~K, $\log g=3.61$, and [Fe/H]$=-0.63$, corresponding to spectral type A1\,IV.
The \textit{X-shooter Spectral Library} (XSL) \citep{Verro2022} reports the following parameter estimates: $T_{\rm eff}=9262$~K, $\log g=3.65$, and [Fe/H]$=-0.4$ \citep{Arentsen2019}.

Measurements of the magnetic field of HD~174240 are compiled in two catalogues.
The \textit{FORS1 catalogue of stellar magnetic field measurements} \citep{Bagnulo2015} includes a series of FORS1/ESO VLT observations that yielded no significant detection ($\langle B_z\rangle < 50$~G, below the $2\sigma$ level).
\textit{A catalogue and statistical analysis of magnetic stars} \citep{Rustem2023} yields a root-mean-square magnetic field of approximately $79 \pm 76$~G, based on $\sim$4 measurements.
These values are inconsistent with the strong, stable fields characteristic of Ap/Bp stars and therefore do not indicate intrinsic magnetically driven coronal activity.

X-ray emission from HD~174240 was detected in the \textit{ROSAT All-Sky Survey} \citep{Schröder2007,Schroeder2008}, with a luminosity of $\log L_\mathrm{X} \approx 29.3~\text{erg~s}^{-1}$.
The authors interpreted this signal as originating from a hidden late-type companion.
According to SRG/eROSITA observations, the X-ray luminosity of the star amounts to $\log L_\mathrm{X} \approx 29.43~\text{erg~s}^{-1}$.

Thus, the combination of spectral properties, the low \ion{Mg}{II}\,$\lambda2800$ index, the absence of an intrinsic magnetic field, and the presence of X-ray emission indicates that HD~174240 does not exhibit intrinsic chromospheric or coronal activity.
The observed X-ray emission is instead attributed to a low-mass, late-type companion.

\subsection{HD~182422} 

HD~182422 is classified as a star of spectral type A1V.
Our derived effective temperature of $T_{\rm eff} = 9362$~K is in good agreement with the value $T_{\rm eff} = 9772$~K reported in the PASTEL catalogue \citep{Soubiran2016}.
The \textit{Gaia} astrometric quality indicator $\mathrm{RUWE} = 1.135$ does not exceed the sky-dependent threshold $\mathrm{RUWE}_{\mathrm{threshold}} = 1.218$, providing no evidence for significant astrometric binarity \citep{Castro-Ginard2024}.

The \textit{ROSAT All-Sky Survey} provides an upper limit on the X-ray luminosity of HD~182422 is $\log L_\mathrm{X} < 29.55~\text{erg~s}^{-1}$ \citep{Berghoefer1996}, while SRG/eROSITA observations yield $\log L_\mathrm{X} = 29.72~\text{erg~s}^{-1}$, which may indicate the presence of a hidden companion.

\subsection{O-type stars}

\begin{figure*}[!t]
\center{\includegraphics[width=0.8\linewidth]{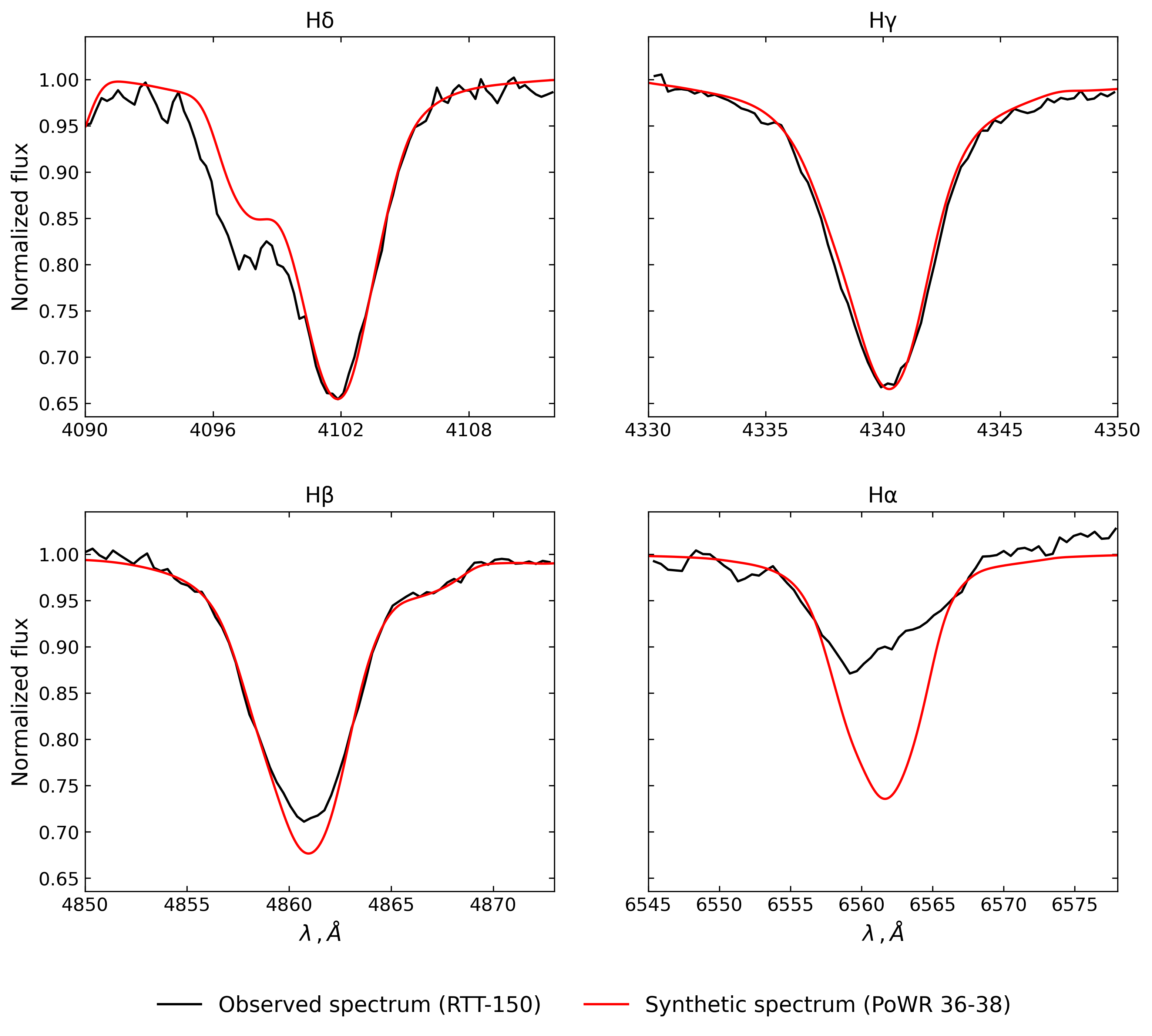}}
\caption{Observed spectrum of HD~157857 compared with a PoWR synthetic spectrum ($T_{\rm eff} = 36\,000$~K, $\log g = 3.8$), which provides the best fit to the observed data.}
\label{Ha_Hb_Hg_HD157857}
\end{figure*}

\begin{figure*}
\center{\includegraphics[width=1.0\linewidth]{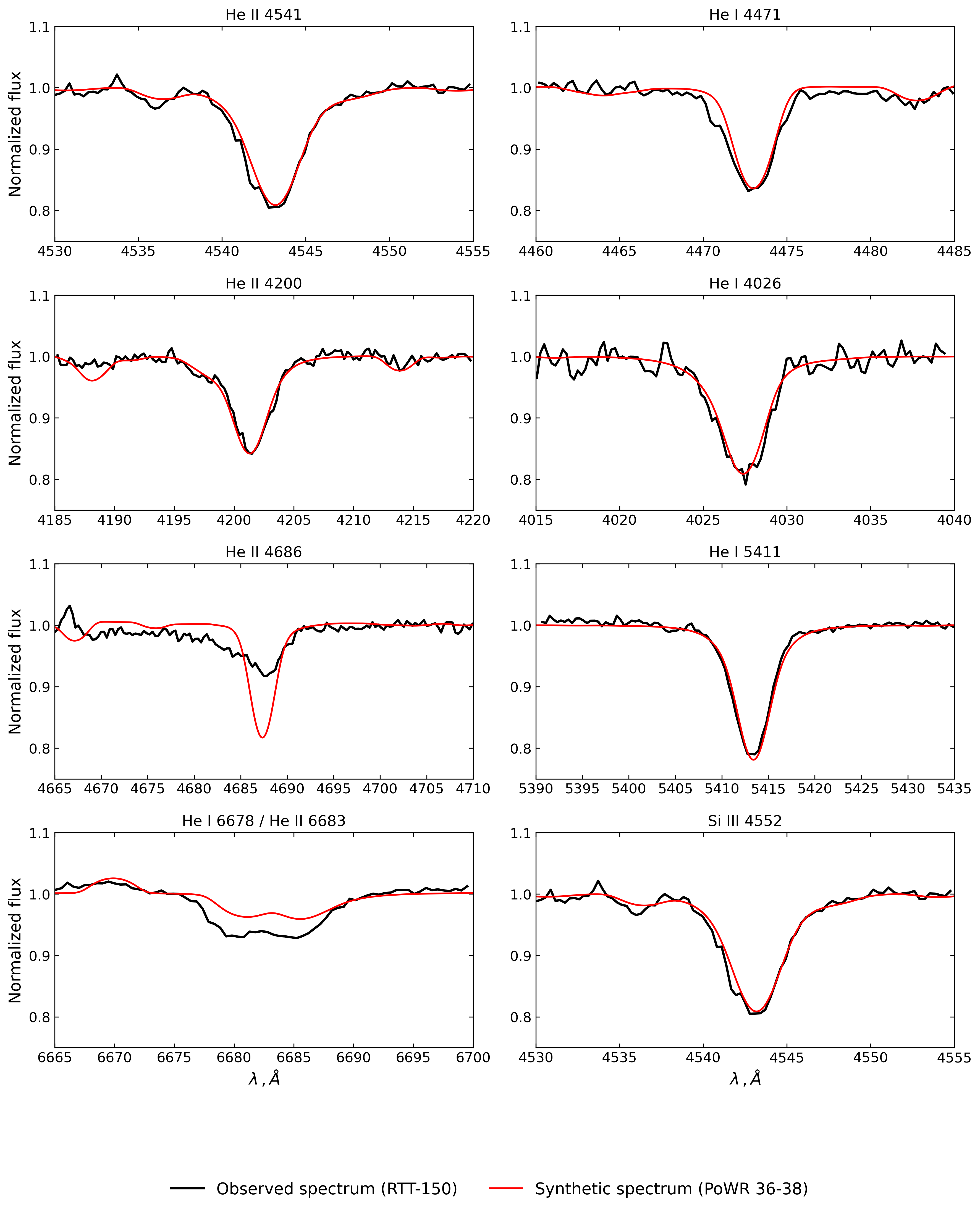}}
\caption{Comparison of the PoWR model with the He\,I, He\,II, and Si\,III lines of HD~157857 (cf. Fig.~\ref{Ha_Hb_Hg_HD157857}).}
\label{He_HD157857}
\end{figure*}

HD~157857, HD~163800, and HD~164438 have been classified within the Galactic O-Star Spectroscopic Survey (GOSSS; \citealt{Sota2011,Sota2014}) as O6.5\,II(f), O7.5\,III((f)), and O9.2\,IV stars, respectively.
\cite{Martins2005} provides calibrations of stellar parameters for Galactic O-type stars across different luminosity classes.
Figures~\ref{Ha_Hb_Hg_HD157857} and~\ref{He_HD157857} show the best-fitting PoWR models computed using the stellar parameters derived with the astroARIADNE code. The models reproduce the hydrogen, helium, and \ion{Si}{III} lines observed in HD~157857. The effective temperatures derived in the present work are in good agreement with those expected from the spectral classifications and the calibrations of \citet{Martins2005}, who employed the CMFGEN code, while the surface gravities are systematically higher, which may be attributed to the use of different stellar atmosphere codes (CMFGEN and PoWR).

A weak longitudinal magnetic field of $\langle B_z \rangle = -110 \pm 46$~G, measured using all available spectral lines, has been detected in HD~157857 \citep{Hubrig2013}.
Weak H$\alpha$ emission is present in the spectrum (Fig.~\ref{Ha_Hb_Hg_HD157857}), indicating the presence of a stellar wind and possibly weak magnetospheric activity.
Based on data from the \textit{Einstein} X-ray Observatory \citep{Chlebowski1989}, only an upper limit on the X-ray luminosity of $\log L_{\mathrm{X}} < 32.76$ could be derived for this star.

HD~163800 is a typical representative of late O7–O8 giants with a moderately strong radiatively driven stellar wind.
Its spectrum shows \ion{N}{III}\,$\lambda\lambda4634$–4642 and \ion{He}{II}\,$\lambda4686$ emission features, characteristic of the ((f)) classification and attributed to recombination emission in the stellar wind.

HD~164438 is a spectroscopic binary system consisting of an O9.2 subgiant and a B5 main-sequence companion \citep{Blex2024}.
The orbital period of the system is $10.2$~d.
The H$\alpha$ line shows emission filling, suggesting a small but non-negligible contribution from the stellar wind.

According to SRG/eROSITA data, all three stars have X-ray luminosities on the order of $\log L_{\mathrm{X}} \approx 31.5$, consistent with standard shock-heating processes in unstable radiatively driven stellar winds \citep{Feldmeier1997}.

\section{Stars with Hidden Companions. Discussion}

To investigate the origin of the X-ray emission, we derived the fundamental parameters of 15 OBA-type stars with X-ray luminosities in the range $\log L_\mathrm{X} = 28.5{-}31.5$, based on SRG/eROSITA data.
We find that the X-ray emission of five OB-type stars originates either in circumstellar envelopes or in collisional processes within stellar magnetospheres.
The presence of stellar winds is supported by the observed composite H$\alpha$ line profiles, which consist of photospheric absorption combined with emission in the line core produced by the wind component in the circumstellar environment.
In eight A-type stars, the H$\alpha$ line profiles are well described by photospheric absorption alone.
Moreover, available studies indicate that these stars do not host strong global magnetic fields.
Consequently, the X-ray emission in these eight objects is attributed to the coronae of hidden cool companions—late-type dwarf stars.

We identify the presence of hidden companions in two stars from the list of ``bona fide'' systems presented by \citet{Schröder2007}.
In the second probabilistic list of binary candidates, one system is confirmed while another is rejected.

For 36~Lyn, the presence of a hidden companion is firmly established, and the companion is classified as an F0\,V star.
For objects of this type, the typical bolometric luminosity is $\log L_{\mathrm{bol}} = 0.89$ \citep{Pecaut2013}.
The resulting ratio $\log(L_\mathrm{X} / L_{\mathrm{bol}}) = -5.14$ is consistent with the unsaturated regime of coronal activity typical of F0\,V stars.
Nevertheless, the possibility that the X-ray emission originates from magnetic activity of the primary star itself cannot be entirely excluded.

For HD~182422, we derive an age of approximately 234~Myr.
The hidden companion is likely of comparable age and therefore belongs to the population of young late-type stars (G--M) that still reside in the saturated regime of coronal activity, for which $\log(L_\mathrm{X} / L_{\mathrm{bol}}) \approx -3$.
Adopting this value allows us to estimate the bolometric luminosity of the companion and hence its spectral type.
The derived value corresponds to a late-type star of approximately K5\,V--K7\,V.

For the remaining stars the spectral type of the
hidden companion may range from F to M. 
The range $\log(L_\mathrm{X}/L_{\mathrm{bol}}) \sim -3$ to $-5$, characteristic of coronally active stars, translates into bolometric luminosities consistent with spectral types from M dwarfs to F-type stars.
As illustrated in Fig.~12 of \cite{Johnstone2021}, at $L_\mathrm{X} \sim 10^{29}$~erg~s$^{-1}$ the allowed mass range spans this entire interval, making an unambiguous determination of the spectral type not possible.

\section{ACKNOWLEDGMENTS}

This study is based on observations with the eROSITA telescope onboard the SRG observatory. The SRG observatory was built by Roskosmos in the interests of the Russian Academy of Sciences represented by the Space Research Institute (IKI) within the framework of the Russian Federal Space Program, with the participation of the Deutsches Zentrum fur Luft- und Raumfahrt (DLR). ¨
The SRG/eROSITA X-ray telescope was built by a consortium of German institutes led by MPE, and supported by DLR. The SRG spacecraft was designed, built, launched, and is operated by the Lavochkin Association and its subcontractors. The science data are downlinked via the Deep Space Network Antennae in Bear Lakes, Ussurijsk, and Baykonur, funded by Roskosmos. The eROSITA data used in this paper were processed with the eSASS software developed by the German eROSITA consortium and the software developed by the Russian SRG/eROSITA consortium.\\

This work has made use of data from the European Space Agency (ESA) mission {\it Gaia} (\url{https://www.cosmos.esa.int/gaia}), processed and analyzed by the Gaia DPAC consortium (\url{https://www.cosmos.esa.int/web/gaia/dpac/consortium}). \\

E.A. Nikolaeva thanks the “Summer School for Astrostatistics in Crete - 2024” for training on the statistical methods used in this paper.

\section{FUNDING}

We are grateful to TUBITAK, the Space Research Institute, the Kazan Federal University, and the Academy of Sciences of Tatarstan for their partial support in using RTT-150 (the 1.5-m Russian–Turkish telescope in Antalya). \\

The work of E.A. Nikolaeva, I.F. Bikmaev, and E.N. Irtuganov was supported by subsidy FZSM-2023-0015 of the Ministry of Education and Science of the Russian Federation granted to the Kazan Federal University for the State assignment in the sphere
of scientific activities. \\

\end{document}